# Soft Sides of Software

Luiz Fernando Capretz
Western University, Department of Electrical and Computer Engineering, London, N6A5B9, Canada
lcapretz@uwo.ca

Faheem Ahmed
Thompson Rivers University, Department of Computing Science, Kamloops, V2C0C8, Canada
fahmed@tru.ca

Fabio Q. B. da Silva
Universidade Federal de Pernambuco, Centro de Informática, Recife, 50740-560, Brazil
fabio@cin.ufpe.br

It is impossible to exclude the human factors from software engineering expertise during software development because software is developed by people and for people. The intangible nature of software has made it a difficult product to successfully create, and a close examination of the reasons for major software system failures shows that several of these reasons eventually boil down to human issues. As software practitioners are immersed in the technological aspect of the product, they can quickly learn lessons from technological failures and readily come up with solutions to avoid them in the future. Nonetheless, they do not learn lessons from the human aspects of software engineering.

Dealing with human errors is much more difficult for developers and often this aspect is overlooked in the development process as software developers move on to problems that they are more comfortable solving. The main reason for the oversight is that human factors are usually related to soft skills, i.e., teamwork, motivation, emotions, commitment, leadership, multi-culturalism, interpersonal skills, etc. Another reason is that there has been a lack of progress in this area since the field of software psychology (a soft side) has not focused on problems arising from human failings to the same extent as we have on strictly technical problems.

A quick search for "human factors" in the IEEE Guide to the Software Engineering Body of Knowledge (SWEBOK) and the ACM/IEEE Curriculum Guidelines for Undergraduate Degree Programs in Computer Science, reveals that the term appears only eight times in each document body. Nevertheless, one prominent sentence does reinforce the importance of the topic: "*Students need to repeatedly see how software engineering is not just about technology.*" However, due to the constraints of most software curricula, the reality indicates that, at best, human factors are squeezed into only one or two HCI courses. SWEBOK-v3.0 suggests that only five hours of studies be given to group dynamics. This is clearly not enough for a topic of such crucial importance. Educators willing to venture into this area face an arduous task if they try to convince their colleagues and software engineering zealots of the importance of soft skills materials.

One factor that may have influenced this lack of attention in the past is that very few researchers and practitioners have explored programming as an individual cognitive activity and not looked at personality traits. Others have touched on team and social perspectives of software engineering and focused on the human aspects in the software development processes. One can find trustworthy materials and heated discussions at the website of the Psychology of Programming Interest Group (www.ppig.org). Nevertheless, studies on human factors to date have only scratched the surface of their impact on the software development process. Occasionally papers have described results obtained by quantitative and qualitative research conducted in this field. Quick searches at Google Scholar have shown a few thousands results for "human factors in software", "psychology of computer programming" and "software psychology", but hundreds of thousands results for "cloud computing" – a much more recent topic in the computing/software arena. Google shows only seven entries for "course in software psychology." Even worse is the fact that this kind of research has had a minimal impact on the daily life of professional software engineers in the last 40 years.

**Pioneers and Late Research on Software Psychology**

Software engineering has come a long way from its defining days of the 1970-s. While it has been excelling in serving diverse requirements of disparate customer bases, ranging from space scientists and weather forecasters to boutiques and retail shop owners, it has also been causing serious concerns due to major system failures caused by software glitches or improper software verification and validation, or human limitations. However, human aspects of software engineering continue to be a neglected research area. Possible reasons for this neglect are: the complex relationships between human psychology and the software development processes, lack of awareness of the impact of human factors in software engineering, and possibly lack of trust in empirical studies on human factors in software engineering. If the *status quo* lasts long, the software engineering discipline may face serious problems.

The importance of the people dimension has been highlighted by thoughtful leaders, like Weinberg [1], Dijkstra [2] and DeMarco [3]. Since the 1970, egoless programming is one of the most cited and most misunderstood concepts in software psychology; it has given rise to a variety of powerful software review techniques. Lately, Weinberg stated that "there is no shortage of evidence that, for example, technical reviews lead to more reliable code produced more cheaply and consistently. And, indeed, more software organizations today regularly use some form of technical review as a standard part of their software processes". This is one of the first pieces of evidence that good software engineering best practices influenced by egoless programming, i.e. reviews and walkthroughs, outlast specific technologies such as old-fashioned CASE tools.

Recently, Cruz et al. [4] conducted an extensive mapping study, in which 19,000 articles published between 1970 and 2010 were retrieved, but only 90 were considered to be representative and relevant to the understanding of the role of individual personality in software engineering. This clearly confirms that despite being a significantly important piece of the software engineering puzzle, the personality factor is still missing when it comes to empirical evidence of realities. Many of the empirical studies to date have revolved around discovering the personality traits of a software engineer while at school or working in the profession. However, little evidence is available on the effectiveness or impact analysis of the personality profile needed in managing a software project, developing group cohesiveness, dealing with individual behavior, conflict management, etc. The micro-level interpretation of software development activities (such as system analysis, design, coding, and testing) demands, in order to effectively carry out the activity, a certain set of abilities from the individuals involved. Determining the best personality traits for these particular roles and the individuals who have these personality traits are concepts that are rarely discussed. Perhaps, that is because determining answers to these questions is not a simple matter.

Undoubtedly, it is important to assign people with particular personality traits to their preferred tasks in a software project; this increases the chances of a successful project outcome [5]. This study tackles a difficult-to-measure aspect of software engineering: that is, how to best choose individuals for the various tasks in a software project. To a certain extent, successful approaches use psychological types to determine who prefers certain software development roles. The study found patterns that link personality traits to role preferences in a software life cycle. Among the various roles, the most preferred ones among the participants are system analyst, software designer, and programmer. In contrast, tester and maintainer happen to be the least popular roles among software engineers. However, that study omits the different characteristics that may be most appropriate for other software occupations, such as project manager, troubleshooter, helpdesk personnel, database administrator, and so forth.

**The Importance of Soft Skills**

When software employers advertise jobs, they divide their wish list into technical and non-technical skill sets. Technical skills are relatively easy to evaluate by looking into academic credentials, certifications, professional experience, etc. On the other hand, the difficulty of assessing non-technical skills – such as interpersonal skills, teamwork, ability to work under strict deadlines, being a fast learner, and open and flexible to change – tend to make these skills overlooked compared to technical skills when employers evaluate candidates for jobs [6]. About 80% of the individuals who fail at work do not fail due to a lack of technical skills but rather because of their inability to relate or communicate well with others in a team [7]. Software development is a collaborative type of work in which solo performers are rare. In this case, an individual who has appropriate academic credentials but is unable to work in a group setting may have a catastrophic effect on the project. However, as we have mentioned, these non-technical skills are difficult to assess at the time of hiring. Similarly, the rapid growth in technology and continuous process improvement are characteristics of software development that make work difficult for someone who has an inability to learn fast or work under constant pressure.

Kappelman et al. [8] provided insights into the diverse and dynamic nature of skills required at different stages of a software engineering career, from new hires to CIOs. They assert that the key to progression to software project management is to hone one's technical and functional-area skills, and that communication skills are critical throughout a software professional career; they advised software professionals to build their people and decision-making skills. Organizations can use these skills to enhance their software-related workforce practices, and software engineers can use them to achieve their personal career objectives and help others do so too. It appears that the time is now ripe to address essential soft skills that complement technical skills.

**Emotions, Culture, and Diversity**

Modern software development is the result of a complex process that involves many stakeholders; these stakeholders include product owners, quality assurance teams, the project manager, and, above all, the software developers. The main difficulty that software project managers face when tasks are assigned is selecting the right people within a team so that the chances of successful completion of the project increases. There is no easy gauge to measure selecting the "right" person for a job because the answer is not deterministic. For example, "motivation" deals with a reason to perform better; however, motivation by itself is often insufficient for achieving goals. Arguably, other human-related factors play an important role in software development. Emotions, moods and feelings in the workplace receive significant attention in management research and organizational psychology. Using biometrics measures, Muller and Fritz [9] show that a wide range of emotions (happiness, frustration, anger, etc.) experienced by software developers is definitely correlated with their progress on the tasks.

Furthermore, software development no longer takes place within one room or building. Instead, it is a global venture performed by development teams composed of individuals scattered across the globe, most of them having different cultural backgrounds. Culture has to do with the way people think, react to events, socialize, and prioritize things, and also the work ethic that they have. How these diverse individuals operate within the development team can present a complex problem to solve; for example, in both the USA and the Netherlands, individualism is very high; whereas, in the cases of China, West Africa, and Indonesia, collectivism is more important in social behavior [10]. Similarly, some cultures are task-oriented instead of relationship-focused. When individuals from these opposite cultures interact to develop a software product, the success of the software project can be relatively difficult to predict, and these opposing factors have the potential to increase project risks. Despite the awareness that these cultural dynamics may contribute to the probable success or failure of the software project, the real issue is that the software industry tends to ignore these dynamics because no one has a clear solution for the problem.

However, the software industry cannot afford to lose potential professionals who may think differently. In terms of software development, better software will result from the combined efforts of a variety of mental processes, experience, and values. Different ways of thinking are important for software engineering, as individuals with different mindsets can make unique contributions during the software development process. More than ever, software engineering needs diversity of thinking because it takes variety to conquer complexity. Binging this to the software context, skills diversity is needed to solve the myriad problems related to software development and maintenance [11]. Since strong teams are the ones made up of balanced perspectives, organizations can benefit from a conscious attempt to diversify the styles of their software engineers. Diversity and variety will enable us to bring a richness of talents and points of view to bear upon the inherent complexity of software systems.

**Teams and Interactions**

Although the research on diversity has led to important results, it has not fully addressed the effects, positive and negative, of having different individuals working and interacting in software development teams. The vast majority of software systems of practical relevance are developed by teams, not individuals, due to their inherent complexity and also to their size and effort needed for their development. When individuals must work in teams, a broader view must be taken: we need to understand how individuals interact and work together in those teams, and this is much more difficult to understand and requires further attention from researchers in behavioral psychology, management science, and empirical software engineering. Consequently, team processes and interactions must be taken into account during team building and throughout the entire lifetime of the team [12]. More recently some embryonic empirical studies have been conducted to exam the impact of personalities in software development teams. Acuna et al. [13] found a positive relationship between some organizational climate factors and satisfaction in software development teams: the teams whose members score highest for the agreeableness personality factor have the highest satisfaction levels, and this has an impact on the software quality. Finally, Yilmaz et al. [14] indicated that effective team structures support teams with higher emotional stability, agreeableness, extroversion, and conscientiousness personality traits. This complex and overlooked research area needs further investigation.

Software is a field of rapid changes: the best technology today becomes obsolete in the near future. If we review the graduate attributes of any of the software engineering programs across the world, life-long learning is one of them. The social and psychological aspects of professional development is linked with rewards. In organizations, where people are provided with learning opportunities and there is a culture that rewards learning, people embrace changes easily. However, the software industry tends to be short-sighted and its primary focus is more on current project success; it usually ignores the capacity building of the individual or team.

It is hoped that our software engineering colleagues will be motivated to conduct more research into the area of software psychology so as to understand more completely the possibilities for increased effectiveness and personal fulfillment among software engineers working alone and in teams.